\documentclass[letter]{aa} 

\usepackage{graphicx}
\usepackage{txfonts}
\usepackage[]{natbib}
\bibpunct{(}{)}{;}{a}{}{,}
\newcommand{\doce}{\mbox{$^{12}$CO}}

\newcommand{\trece}{\mbox{$^{13}$CO}}

\newcommand{\jtd}{\mbox{$J$=3$-$2}}

\newcommand{\kms}{\mbox{km\,s$^{-1}$}}

\newcommand{\gsim}{\raisebox{-.4ex}{$\stackrel{\sf >}{\scriptstyle\sf \sim}$}}

\newcommand{\secp}{\mbox{\rlap{.}$''$}}
\newcommand{\degp}{\mbox{\rlap{.}$^\circ$}}
\begin{document}

   \title{High-resolution observations of the symbiotic system R Aqr}
   
   \subtitle{Direct imaging of the gravitational effects of the
     secondary on the stellar wind}

   \author{V. Bujarrabal
          \inst{1}
          \and J. Alcolea\inst{2}
          \and J. Miko\l ajewska\inst{3}
          \and A. Castro-Carrizo\inst{4}  \and
S. Ramstedt\inst{5} }

   \institute{             Observatorio Astron\'omico Nacional (OAN-IGN),
              Apartado 112, E-28803 Alcal\'a de Henares, Spain\\
              \email{v.bujarrabal@oan.es}
              \and
             Observatorio Astron\'omico Nacional (OAN-IGN),
             C/ Alfonso XII, 3, E-28014 Madrid, Spain
              \and
              Nicolaus Copernicus Astronomical Center, Polish Academy
              of Sciences, ul.\ Bartycka 18, PL-00-716 Warsaw, Poland
\and 
 Institut de Radioastronomie Millim\'etrique, 300 rue de la Piscine,
 38406, Saint Martin d'H\`eres, France
\and 
 Department of Physics and Astronomy, Uppsala University, Box 516,
 75120 Uppsala, Sweden  
   }

   \date{submitted 13 June 2018, accepted 11 July 2018}

  \abstract {We have observed the symbiotic stellar system R Aqr,
    aiming to describe the gravitational interaction between the white
    dwarf (WD) and the wind from the Mira star, the key phenomenon
    driving the symbiotic activity and the formation of nebulae in such
    systems.  We present high-resolution ALMA maps of the \doce\ and
    \trece\ \jtd\ lines, the 0.9 mm continuum distribution, and some
    high-excitation molecular lines in R Aqr. The maps, which have
    resolutions ranging between 40 milliarcsecond (mas) and less than
    20 mas probe the circumstellar regions at suborbital scales as the
    distance between the stars is $\sim$ 40 mas.  Our observations show
    the gravitational effects of the secondary on the stellar wind. The
    AGB star was identified in our maps from the continuum and
    molecular line data, and we estimated the probable position of the
    secondary from a new estimation of the orbital parameters. The
    (preliminary) comparison of our maps with theoretical predictions
    is surprisingly satisfactory and the main expected gravitational
    effects are directly mapped for the first time. We find a strong
    focusing in the equatorial plane of the resulting wind, which shows
    two plumes in opposite directions that have different velocities
    and very probably correspond to the expected double spiral due to
    the interaction. Our continuum maps show the very inner regions of
    the nascent bipolar jets, at scales of some AU. Continuum maps
    obtained with the highest resolution show the presence of a clump
    that very probably corresponds to the emission of the ionized
    surroundings of the WD and of a bridge of material joining both
    stars, which is likely material flowing from the AGB primary to the
    accretion disk around the WD secondary.  }
  
   \keywords{stars: AGB and post-AGB -- circumstellar matter --  binaries: close --
     binaries: symbiotic  --  stars: individual: R Aqr}

   \maketitle
%

\section{Introduction}

Symbiotic stellar systems (SSs) are interacting binaries consisting of
an evolved cool giant and a compact companion, usually an AGB star and
a white dwarf (WD).  In classical SSs, the interaction is very
strong; there are copious mass transfer, equatorial flows, and ejection of
fast bipolar jets.
The relevance in the SS activity of the gravitational interaction
between the wind from the primary and the compact secondary is
stressed in recent 3D simulations, see \citet{mohamedp12},
\citet{valborro17}, and \citet{saladino18}. Those calculations revealed a
new mass-transfer mode called wind Roche-lobe overflow (WRLOF), which
happens when grains form in the AGB circumstellar envelope beyond the
Roche lobe and the wind velocity is still moderate when it reaches the
surroundings of the secondary, significantly reinforcing the
companion-wind interaction. The resulting outflows are strongly
focused toward the binary orbital plane and mass-transfer and accretion
rates are at least an order of magnitude higher than previously
predicted.

R Aqr is the best studied SS. The primary is a bright Mira-type
variable and the companion is a WD. The two-arcminute-wide nebula is
composed of an equatorial structure elongated in the east-west
direction and a precessing jet (with position angle, PA, ranging
between 10$^\circ$ and 45$^\circ$) powered by the accretion onto the
WD; see \cite{solfu85}, \cite{melnikov18}, and references therein. The
orbital period of the binary system is long, $\sim$ 43.6 yr, and the
orbital plane is roughly perpendicular to the plane of the sky and
projected in the east-west direction \citep{gromik09}.  Recent Very
Large Telescope (VLT) imaging by \citet{schmid17} resolved both stars,
which were found to be separated by $\sim$ 45 milliarcsecond (mas). The
photospheric diameter of the AGB star, $\sim$ 10-20 mas, was measured
from IR interferometry by \cite{ragland08} and \cite{wittk16}.
Molecular emission is very rarely observed in SSs, probably because of
photodissociation by the UV emission of the WD and its surroundings or
to strong disruption of the shells, except from regions very close to
the AGB. However, R Aqr has been detected in SiO, H$_2$O, and CO
emission and is relatively well studied in molecular lines
\citep[][]{bujetal10}. Parallax measurements from SiO Very Long
Baseline Interferometry (VLBI) data indicate a distance of 218 pc
\citep[][]{min14}, although the distance from the GAIA parallax is 320
pc; the origin of such a discrepancy is unknown. Both measurements are
subject to uncertainties, because the stellar diameter is larger
than the measured parallax and the SiO emission is still wider and less
uniform than the stellar disk.

Predictions of WRLOF models agree with the observed large-scale nebular
structure in SSs and the requirements to explain their
activity.  However, there is no direct observational information on
how the gravitational effects of the secondary on the stellar wind
take place. We present ALMA maps of radio continuum and
molecular lines in R Aqr that show very clearly those 
effects at orbital and suborbital scales of $\sim$ 10-40 mas.

\section{Observations}

Observations were performed with ALMA on November 21 and 23,
2017, for three tracks of 1.3 h each.  Data were obtained with 47-48
antennas, with baselines ranging from 92 m to 8.5 km.  The correlator
was set to observe with four spectral windows, centered at frequencies
330583, 331295, 343495, and 345791 MHz, and with spectral resolutions
0.24, 0.98, 0.98 and 0.12 MHz, eventually smoothed in the final
maps. The quasar J2348-1631 (1\degp 6 away from R Aqr) was the phase
calibrator, and J0006-0623 was the bandpass and flux calibrator: a
flux reference of 1.58/1.65 Jy (at the lowest/highest frequencies) was
adopted for the three tracks. Differences in the phase calibrator flux
of 5\% between consecutive tracks were found, which is a limit to the
flux uncertainty.
The data calibration was performed using the ALMA pipeline delivered
with the CASA software.

For image cleaning, we used the Hogbom method and the Briggs weighting
scheme with a robust value of 1, resulting in maps with half-power beam
width (HPBW) $\sim$ 40$\times$35 mas, see Figs.\ 1 and 3. The
  half-power field of view and the maximum recoverable scales are
  $\sim$18$''$ and $\sim$2$''$,   respectively.  To better investigate the
compact continuum clump, we also produced images of less sensitivity
but higher spatial resolution using only data from baselines longer
than 2.5 km and uniform weighting, which resulted in a beam of
17$\times$27 mas. The distribution of the clean components (i.e. where
the flux is deduced to come from) in this map was analyzed by means of
a yet higher resolution image, by imposing a circular restoring beam of
10 mas (red contours in Fig.\ 2).  See App.\ A for a more detailed
description of our continuum mapping.

\section{Results and conclusions}

\subsection{Summary of observational results}

We have detected several molecular lines as well as the continuum
emission at $\lambda$ = 0.9 mm from R Aqr. As mentioned, continuum maps
were obtained weighting the visibilities in two different
ways. Together with a more conservative standard procedure that leads
to a resolution of about 30$\times$40 mas, we also used
a weighting that favors long baselines, leading to a
resolution of $\sim$ 10-20 mas. See Sect.\ 2 and Appendix A and Figs.\ 1 and 2.
All our maps are centered on the continuum emission centroid
(ICRS coordinates R.A.: 23:43:49.4962, dec.: --15:17:04.72),  to
  which  the offsets given in this letter always refer.

\vspace{-0cm}
\begin{table}[t]
\begin{center}                                          
\caption{Main line parameters derived for high-excitation lines and
  $^{29}$SiO absorption. We always give the R.A.\ and dec.\ offsets 
  with respect to the total continuum centroid: 23:43:49.4962,
  --15:17:0.4.72.}

\scriptsize
\vspace{-.3cm}
\begin{tabular}{|l|cc|cc|}
\hline
molecular line & total flux & peak brightness & $\Delta$(R.A.) & $\Delta$(dec) \\
      & mJy$\times$km/s & mJy/beam$\times$km/s & mas       &  mas   \\
\hline
H$_2$O  $\nu_2$=2 3(2,1)--4(1,4) & 370$\pm$15 &  174$\pm$7 & 11$\pm$3 &
3$\pm$3 \\  
Si$^{17}$O $v$=1 $J$=8--7 & 100$\pm$6  &  53$\pm$4 & 10$\pm$3 & 2$\pm$3 \\
CO  $v$=1 \jtd\ & 310$\pm$7&  132$\pm$5 & 11$\pm$3 & 3$\pm$3  \\
$^{29}$SiO $v$=0 $J$=8--7 (abs.) & --53$\pm$4   &  --48$\pm$4 &
9$\pm$4  &  0$\pm$4 \\  
SO $^3\Sigma$ $v$=1 9(8)--8(7) & 307$\pm$8 & 130$\pm$5 & 14$\pm$3 &
5$\pm$3 \\ 
\hline
\end{tabular}
\end{center}
\end{table}

\begin{figure}
    \hspace{.4cm}
     \includegraphics[width=7.9cm]{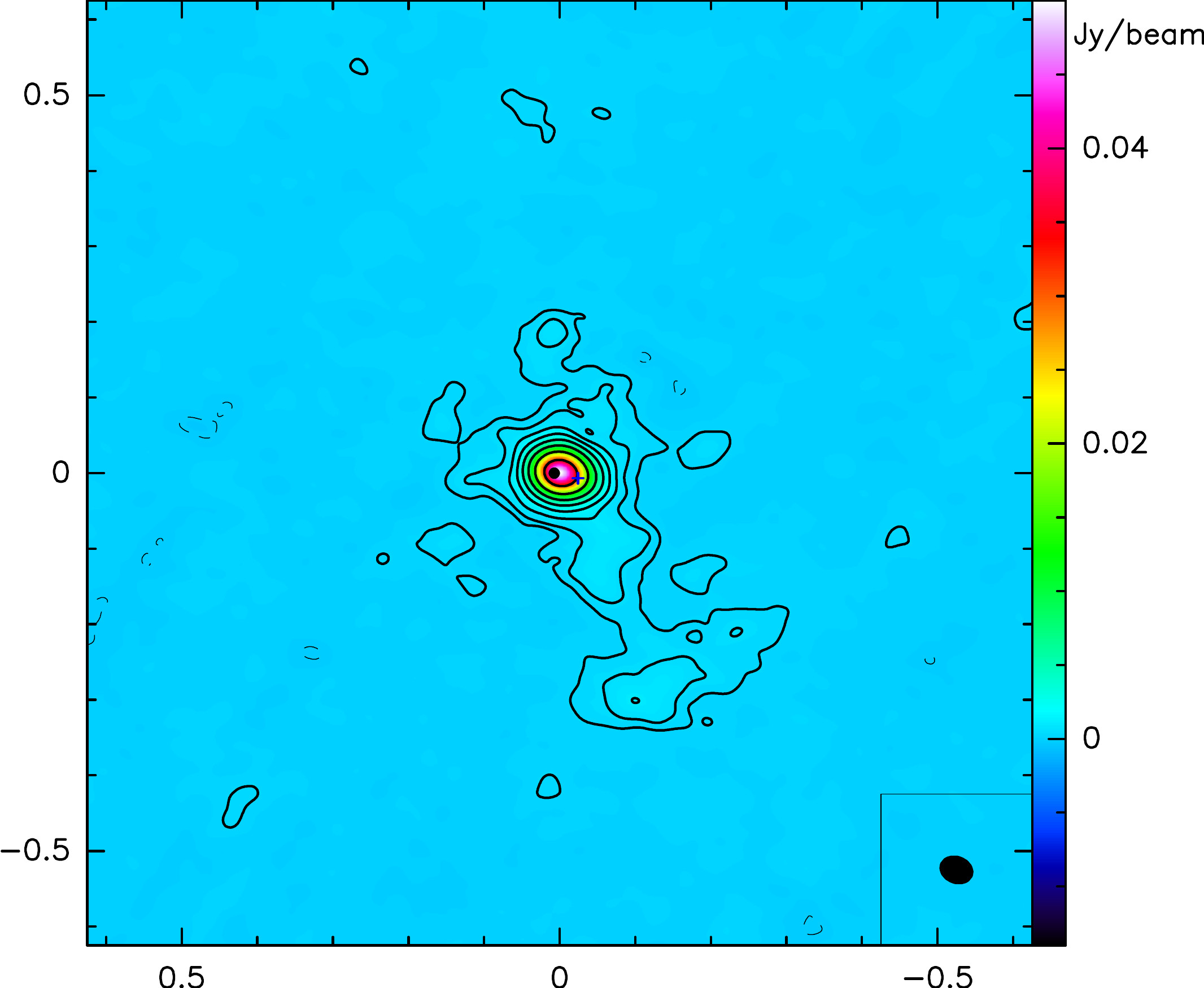}
\caption{ALMA map of the 0.9 mm continuum in R Aqr. The level spacing is
  logarithmic, with first contour at 0.25 mJy/beam and a jump of a
  factor 2; dashed contours represent negative values. The map center
  is the continuum centroid and the positions of the two stars are
  indicated. See the HPBW in the inset.
      }
      \label{}
\end{figure}

\begin{figure}
    \hspace{.5cm}
     \includegraphics[width=6.6cm]{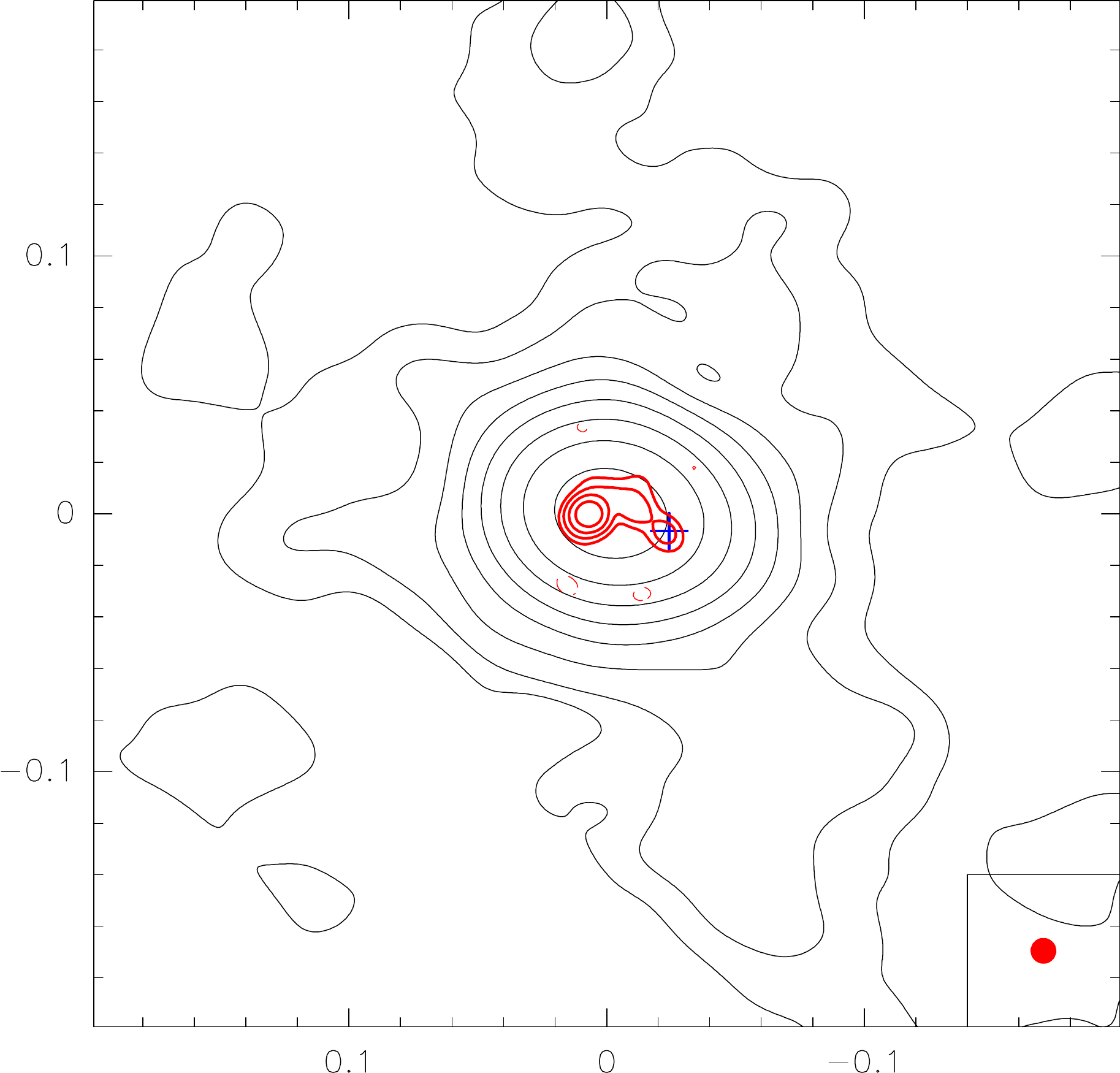}
      \caption{High-resolution continuum map
          obtained using a 10 mas 
        restoring beam (Sect.\ 2), red contours. The scale is logarithmic:
 the first contour is 1.5 
        mJy/beam and the jump is a factor 2; the dashed contours represent
        negative values. The expected position of the WD is
        shown with error bars (blue cross, see Sect.\ 3.2 and App.\ B);
        the Mira is coincident with the maximum of the high-resolution
        map. The central part of our standard continuum map (Fig.\ 1)
        is reproduced in black contours.}
      \label{}
\end{figure}

\begin{figure*}
     \hspace{-.cm}
     \includegraphics[width=17.8cm]{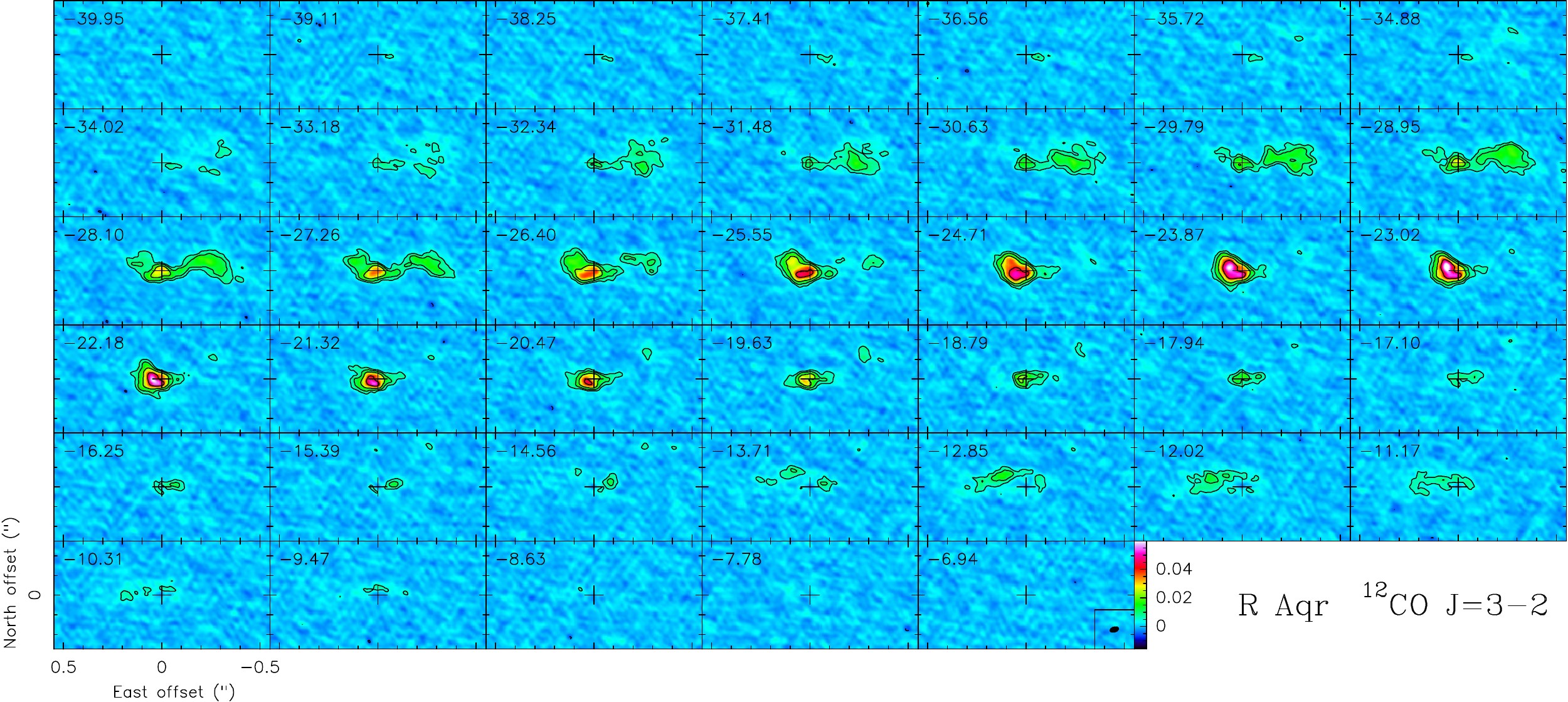}
      \caption{ALMA maps per velocity channel of \doce\ \jtd\ emission
        in R Aqr;
        see the {\em LSR} velocities in the upper left
        corners.  The  center is the centroid of the continuum, whose
        image has been subtracted. The contours are logarithmic
        with a jump of a factor 2 and a first level of $\pm$5 mJy
        (equal to 6.2 times the rms and equivalent to 34.8 K); dashed
        contours represent negative values.  The HPBW is
        shown in the inset, last panel.}
         \label{}
\end{figure*}

The maximum in the high-resolution continuum map is placed (with
respect to the continuum emission centroid) at $\Delta$(R.A.) = +8 and
$\Delta$(dec) = --0.4 mas; i.e.\ the continuum peak coordinates are
ICRS R.A.\ 23:43:49.49657, dec.\ --15:17:04.7204.  As we see in the following, that
maximum is expected to represent the Mira position. By fitting in the
$uv$ plane a disk-like distribution, i.e.\ uniform and round, we
find a 15-mas-wide disk with a center placed at +4.5 and 0 mas. In our
case, the astrometry uncertainties
are dominated by the phase stability (see the ALMA Cycle
5 Technical Handbook); we thus expect absolute errors $\sim$ 3 mas, which
are sufficient for our purposes (Sect.\ 3.3).  A smaller uncertainty, at least in
relative positions, can probably be reached after a deeper data
analysis.

We also detected several molecular lines: the intense \doce\ and
\trece\ \jtd, and, among the weaker lines, H$_2$O $\nu_2$=2
3(2,1)-4(1,4) (331.123730 GHz), Si$^{17}$O $v$=1 $J$=8--7 (332.021994
GHz), \doce\ $v$=1 \jtd\ (342.647636 GHz), $^{29}$SiO $v$=0 $J$=8--7
(342.980847 GHz) and SO $^3\Sigma$ $v$=1 9(8)--8(7) (343828.513 GHz).
The $^{29}$SiO $v$=0 line shows blueshifted absorption against the
stellar continuum (in the range between --29 \kms\ and --33 \kms\ {\em
  LSR}), very probably owing to gas in expansion in front of the
star. The high-excitation ($v$ $>$ 0) lines and the absorption feature
show compact, barely resolved distributions. In Table 1, we give the
main properties of those lines, including the centroids of their
distributions, which are necessary to identify the AGB star
(Sect.\ 3.2), and 1$\sigma$ uncertainties.  Only the CO $v$=0 lines,
mainly the \doce\ line, show a significant extent \gsim\ 0\secp 5
(Sect.\ 3.4).  Recent ALMA maps by \cite{rams18}, which have a
resolution of 0\secp 5, detected clumps at $\Delta$(R.A.) $\sim$
$\pm$0\secp 8, also present in our maps but with a relatively
poor signal-to-noise ratio (S/N).

\subsection{Position of the binary system}

The high-excitation lines detected in our data are expected to come
from the close surroundings of the AGB star and to be good tracers of
its position, as found from VLBI
measurements of R Aqr and ALMA maps of other AGBs; see \citet{min14},
\citet{decin18}, etc.  The absorption in the $^{29}$SiO $v$=0 $J$=8-7
line must represent absorption by inner shells just in front of the
star and should also be a very good tracer of its position.  As we have
seen, see Sect.\ 3.1 and Table 1, all our high-excitation lines and the
$^{29}$SiO $v$=0 absorption show indeed compact images, whose
centroids are practically coincident within the uncertainties and given
the extents of the observed distributions ($\sim$ 35-55 mas). Their
positions can also be considered coincident with the continuum peak
detected with the highest resolution (Sects.\ 3.1, 3.3); the
coincidence between the $^{29}$SiO absorption and the continuum maximum
is particularly good. The differences are significantly smaller than
the expected diameter of the star, $\sim$ 10-20 mas. In any case, the
line emission centroids tend to be shifted by about +3 mas in
R.A.\ with respect to the position obtained from the continuum
(Sect.\ 3.1). It is difficult to discern which of the methods traces
the Mira position more accurately: the measured continuum centers could
be shifted westward, because of contamination from emission from the
extended continuum or irradiation of the primary surface, and the line
emission, even if it is compact, could be slightly shifted eastward
because of the molecule emission suppression observed clearly in
\doce\ (Sect.\ 3.4). Needless to say, it is possible that the
photospheric surface and nearby surroundings are not uniform, which
would not allow comparisons at scales much smaller than $\sim$ 10
mas. We therefore conclude that the continuum peak position, namely
ICRS R.A.\ 23:43:49.49657, dec.\ --15:17:04.7204, gives the AGB
photosphere centroid with an accuracy of $\pm$3 mas.  We have checked
that these coordinates are fully compatible with the GAIA DR2 data. We
recall that the AGB star position is in any case not coincident with
the centroid of the total continuum emission (Sect.\ 2), which is taken
as the center in all maps presented here.

As mentioned before, the two stars were imaged in 2014.9 by
\citet{schmid17}. Our ALMA observations were obtained three years
later, and a moderate, but non-negligible change in the relative
positions is expected. We estimated that change by adapting the
orbital parameters determined by \citet{gromik09} to the measurement
by Schmidt et al. The derivation of the new orbit parameters is
presented in App.\ B; we plan to widely discuss these in a future
paper. Our conclusion is that in 2017 the secondary was placed at
about --31 mas in R.A.\ and --7 mas in declination, with respect to
the Mira star, with an uncertainty of about $\pm$ 7 mas; we note that the
secondary is approaching us.  The positions of both stars are
shown in some of our figures.

\subsection{Continuum 0.9 mm maps}

Our 0.9 mm continuum map is shown in Fig.\ 1; the beam size at half
maximum (FWHM) is $\sim$ 30$\times$40 mas.  The expected positions of
both stars are also given in Fig.\ 1. The AGB primary is
  represented by the black dot, whose width is roughly equal to the
AGB photospheric disk, and the blue cross gives the position of the WD
with the uncertainties; the stellar locations are discussed in the
previous subsection and Apps.\ A, B. As we can see, the main continuum
component is very compact, slightly elongated in the east-west
direction, roughly in the direction of the apparent orbit shape
(see App.\ B).  In addition to the prominent maximum, there is a
low-brightness component elongated in the direction of the large-scale
jet (Sect.\ 1). This component is very probably the base of the bipolar
jets, which we detect down to scales of 30 mas, some AU. The total flux
is $\sim$ 100 mJy, of which $\sim$ 70 mJy comes from the central
condensation; this value is only somewhat larger than the expected flux
coming from the primary photosphere, $\sim$ 35 mJy (for the
photospheric size discussed in Sect.\ 1). The photosphere represents,
therefore, an important component of the total continuum emission at
0.9 mm, with significant contributions from nearby stellar
surroundings and the jet.

To better show the structure of the central continuum, we performed
additional higher resolution maps, selecting only the longest
baselines (Sects.\ 2, 3.2, and App.\ A; see results in Fig.\ 2). This
procedure yields a very high resolution, but also a significant amount
of lost flux and a worse S/N; this was possible only because of the
very high dynamic range of the continuum observations.  We can see
that the jet is resolved out and only the intense compact component is
recovered.
This compact region is composed of a maximum placed on the expected
AGB star position plus an extension in the direction of the companion
(Sect.\ 3.2, Apps.\ A, B). Remarkably, there is a secondary clump
almost exactly coincident with the WD position predicted from our
new orbit determination (Sect.\ 3.2) and a third component joining
both maxima. The intensity and spectral index of these components
support our identifications (App.\ A). We think that this
high-resolution continuum map is actually detecting both stars (or
emission coming from their close surroundings) and the transfer of
material from the primary to the secondary.

\subsection{\doce\  \jtd\ maps}

\doce\ and \trece\ \jtd\ are the only lines that show a significant
extent in our maps, particularly the \doce\ line discussed here. The
\doce\ maps per velocity are shown in Fig.\ 3.  In Fig.\ 4, we present
the map of the velocity-integrated flux; the expected positions of the
two stars are also shown, as derived in Sect.\ 3.2. We also show the
position-velocity diagram found for a central cut in the east-west
direction. In our maps, there is a compact central component plus two
plumes of emitting gas at relatively negative and positive velocities,
extending to the west and to the east, respectively, occupying in total
about 0\secp 7 ($\sim$ 200 AU, 3 10$^{15}$ cm).  We note that the
\doce\ \jtd\ maximum is not placed on the maximum found for the other
lines (including \trece\ \jtd) and for the continuum, but eastward by
20-30 mas. We suggest that this is due to the effects of
photodissociation or shell disruption, which tend to suppress molecular
line emission in SSs (Sect.\ 1) and  must be more
important in regions closer to the companion.

The emission distribution, extending in the east-west direction,
  is similar to the expected image of the two spiral arms predicted by
  WRLOF models; see for example Figs.\ 1, 2 and 3 of \cite{mohamedp12}, model
  M1 (in theoretical papers the velocity is often represented
  in a comoving frame). We recall that the orbit and equatorial planes
are almost seen edge-on with the north pole slightly pointing toward
us, the projected direction of the orbit is roughly in the
east-west direction or slightly inclined (PA \gsim\ 90$^\circ$), and
the secondary is approaching us; see Sect.\ 3.2 and App.\ B. The
CO structure, clearly elongated in the direction of the projected
orbit, obviously corresponds to mass ejection strongly focused in the
orbital plane.  The shape and velocity of the plumes are similar
  to those expected for the projection in the plane of the sky of the
  spiral-like arms.  The westward plume shows a negative velocity shift
  with respect to the systemic velocity of $\sim$ $-$10 \kms, which is
  compatible with expectations: we know that the secondary is moving
  with velocities of this order \citep[from the spectroscopic stellar
    data and stellar dynamics in][]{gromik09} and that it efficiently
  drags the nearby gas (from hydrodynamical calculations). Up to a
  R.A.\ offset of $-$0\secp 2, it shows a slightly curved
  shape with concavity pointing northward, which is the expected
  projected shape of the inner spiral arm that is now being pushed by
  the companion. The outer clump at offset $\sim$ $-$0\secp 3 seems to
  represent the second spiral arm, which must also move toward us but
  at a slightly lower velocity. Finally, emission at relatively
  positive velocities (between $-$14 and $-$9 \kms\ {\em LSR}, Figs.\ 3
  and 4) mostly comes from a plume placed eastward and northward,
  exactly as expected for material accelerated by the passage of the
  secondary about half an orbit ago and that, therefore, moves away
  from us.
  
\begin{figure}
    \hspace{.4cm}
     \includegraphics[width=7.9cm]{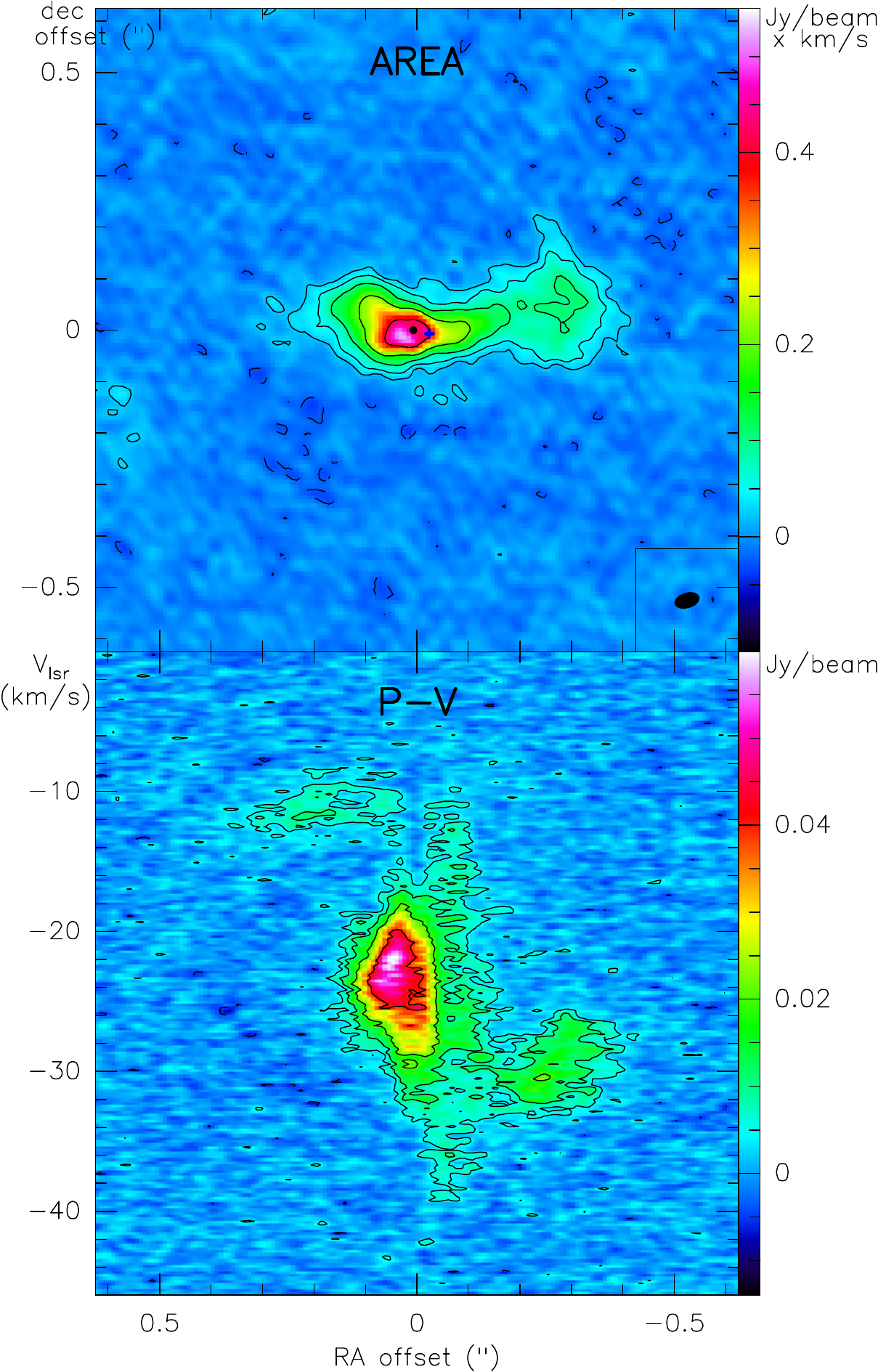}
      \caption{{\em Top:} Map of the \doce\ \jtd\ velocity-integrated brightness in
        R Aqr. Contours are logarithmic, the first contour is $\pm$
        0.025 Jy/beam$\times$km/s and the jump is a factor 2. Dashed
        contours represent negative values.
{\em Bottom:} Position-velocity diagram for an east-west central 
cut.
      }
      \label{}
\end{figure}

The whole set of molecular line data and a quantitative comparison
  with model predictions will be presented in a forthcoming paper.  We
think that our CO maps directly show the very strong
gravitational effects of the WD secondary on the circumstellar wind
leaving the AGB primary, including strong confinement to the equatorial
plane and the formation of a double spiral extending significantly
beyond the orbit.

\begin{acknowledgements}
This work has been supported by the Spanish MINECO, grant
AYA2016-78994-P, and by the National Science Centre, Poland,
grant OPUS 2017/27/B/ST9/01940. This paper makes use of the following ALMA data:
ADS/JAO.ALMA\#2017.1.00363.S. ALMA is a partnership of ESO (representing
its member more states), NSF (USA) and NINS (Japan), together with NRC
(Canada), MOST and ASIAA (Taiwan), and KASI (Republic of Korea), in
cooperation with the Republic of Chile. The Joint ALMA Observatory is
operated by ESO, AUI/NRAO and NAOJ. 
\end{acknowledgements}



\appendix

\section{High-resolution continuum maps}

For the continuum images we only used the two spectral windows with
the highest bandwidths in both receiver side-bands (centered at
331.295 and 343.495 GHz). The data were first mapped seeking for
spectral lines that were flagged out. Line-free data from the two
side-bands were combined and then imaged and cleaned using CASA. As
mentioned, we first used a robustness factor of 1, which resulted in a
clean/restoring beam of 30$\times$40 mas at PA 69$^\circ$ (Sect.\ 2,
Fig.\ 1).  To investigate the strong compact component, we also
produced images using only data from baselines of 2.5 km length and
above and adopting a robustness factor of --2 (equivalent to uniform
weighting). In this way we filter any contribution from structures
with sizes \gsim\ 70 mas out, but reach a higher resolution.  The
final HPBW is 17$\times$27 mas (at PA 51$^\circ$), with an rms of
220\,$\mu$Jy\,beam$^{-1}$; the resulting map is shown in Fig.\,A.1.
The emission appears clearly resolved with a strong central component
and a curved extension first to the west and then to the south; the
jet emission is resolved out.  The peak and total fluxes in this map
are 28\,mJy\,beam$^{-1}$ and 42\,mJy respectively; about 60\% of the
flux is recovered.  Given our limited angular resolution, we analyzed
the distribution of clean components (which represent where in our
maps the emission is deduced to come from), by producing a yet higher
resolution version of the map with a circular restoring beam of 10 mas
HPBW (see Fig.\,2). These procedures are often applied in
radiointerferometry and are justified as far as the restoring beam is
significantly larger than the clean beam size divided by the S/N or,
more restrictively, divided by the intensity ratio of the
significant clean components to those in adjacent regions. We have
checked that the emission from, for instance, the southern, 10-mas wide
area between the maxima outside the lowest level in Fig.\ B.1 is 10-30
times weaker that the maxima themselves.  Continuum emission appears
clearly separated in three locations: from left to right, components A,
B, and C (Fig.\ B.1), which show peak (total) fluxes of 23 (31), 6
(10) and 5\,mJy\,beam$^{-1}$ (5\,mJy), respectively, and an
rms of 250\,$\mu$Jy\,beam$^{-1}$.  Comparing the fluxes obtained
at 331 and 343 GHz separately, we estimated the spectral index of
these three emitting spots; we found values of about 1.9, 3.1, and 1.0 for
components A, B, and C, respectively.

Component A is coincident in position with the emission from
vibrationally excited lines and the absorption feature of $^{29}$SiO
(Sect.\ 3), suggesting that this is indeed the emission from the Mira
component in the system. We checked that this position is also
coincident within the uncertainties with the GAIA coordinates for R
Aqr, which are expected to give the coordinates of the bright primary
(after correcting for proper movements). Its total flux (31 mJy) and
peak brightness are compatible with those expected from the Mira (for a
temperature of $\sim$ 2650 K, Sect.\ 3) and its spectral index
strongly suggests optically thick photospheric emission.  We estimated
the position of the WD companion for the epoch of our ALMA observations
(2017.9), see Sect.\ 3 and App.\ B, which is found to be coincident
with our C component within the errors. This coincidence and the
measured spectral index, which is compatible with emission from the
ionized surroundings of the WD, strongly suggest that this emission 
points to the location of the companion.  Under the assumption that
continuum components A and C identify the primary and the companion in
R Aqr, then the intermediate component B, which shows a spectral index
compatible with dust emission, would be the first detection of mass
transfer between the stars in a SS.

\begin{figure}
    \hspace{.8cm}
     \includegraphics[width=6.7cm]{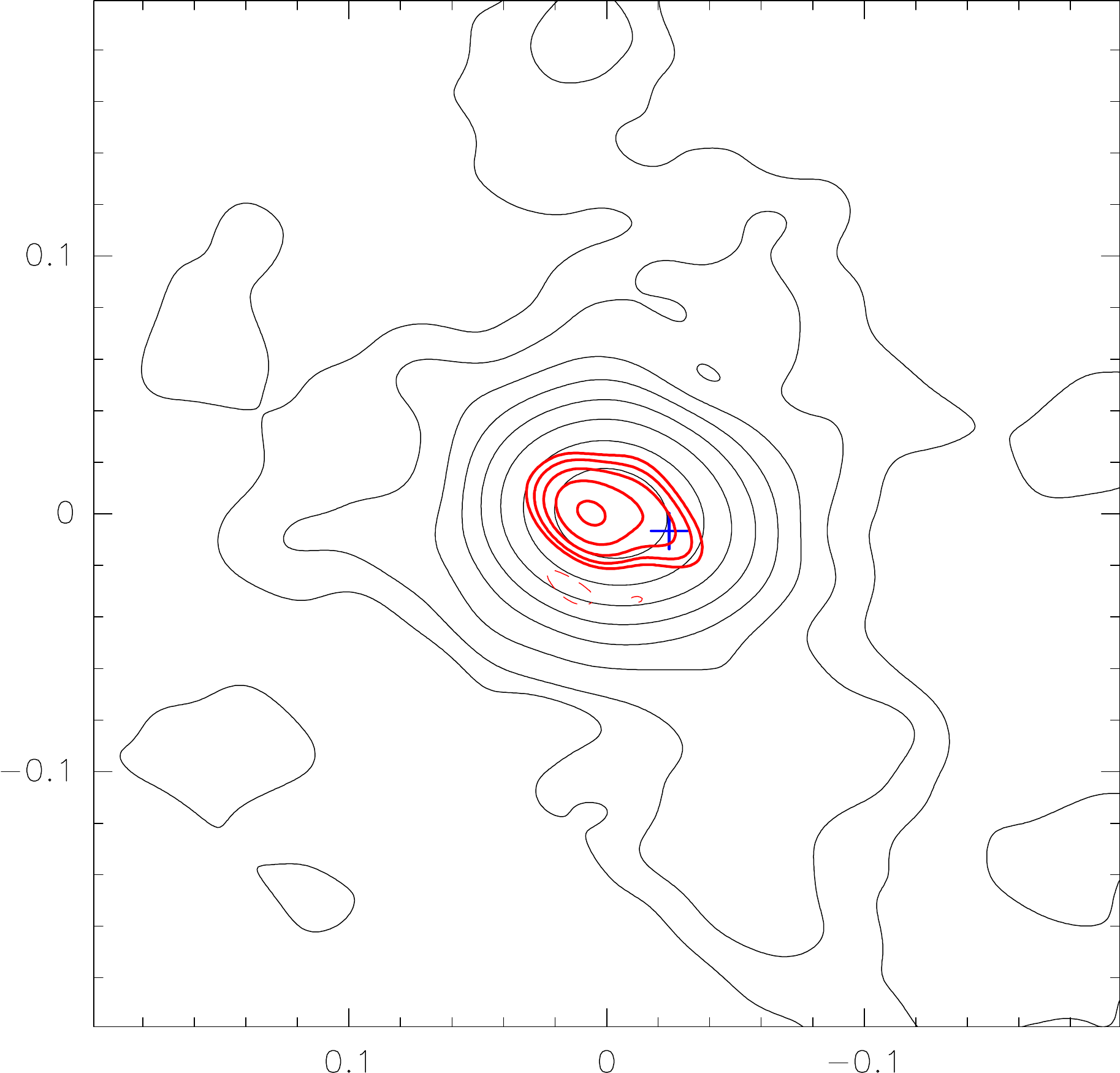}
      \caption{Same as Fig.\ 2, same scales and units, but showing the
        continuum map obtained after 
        increasing the angular resolution and using the
        original clean
        beam (17$\times$27 mas).}
      \label{}
\end{figure}

\section{New orbital parameters determined accounting for
  recent stellar astrometry}

\begin{figure}
    \hspace{1.1cm}
     \includegraphics[width=6.4cm]{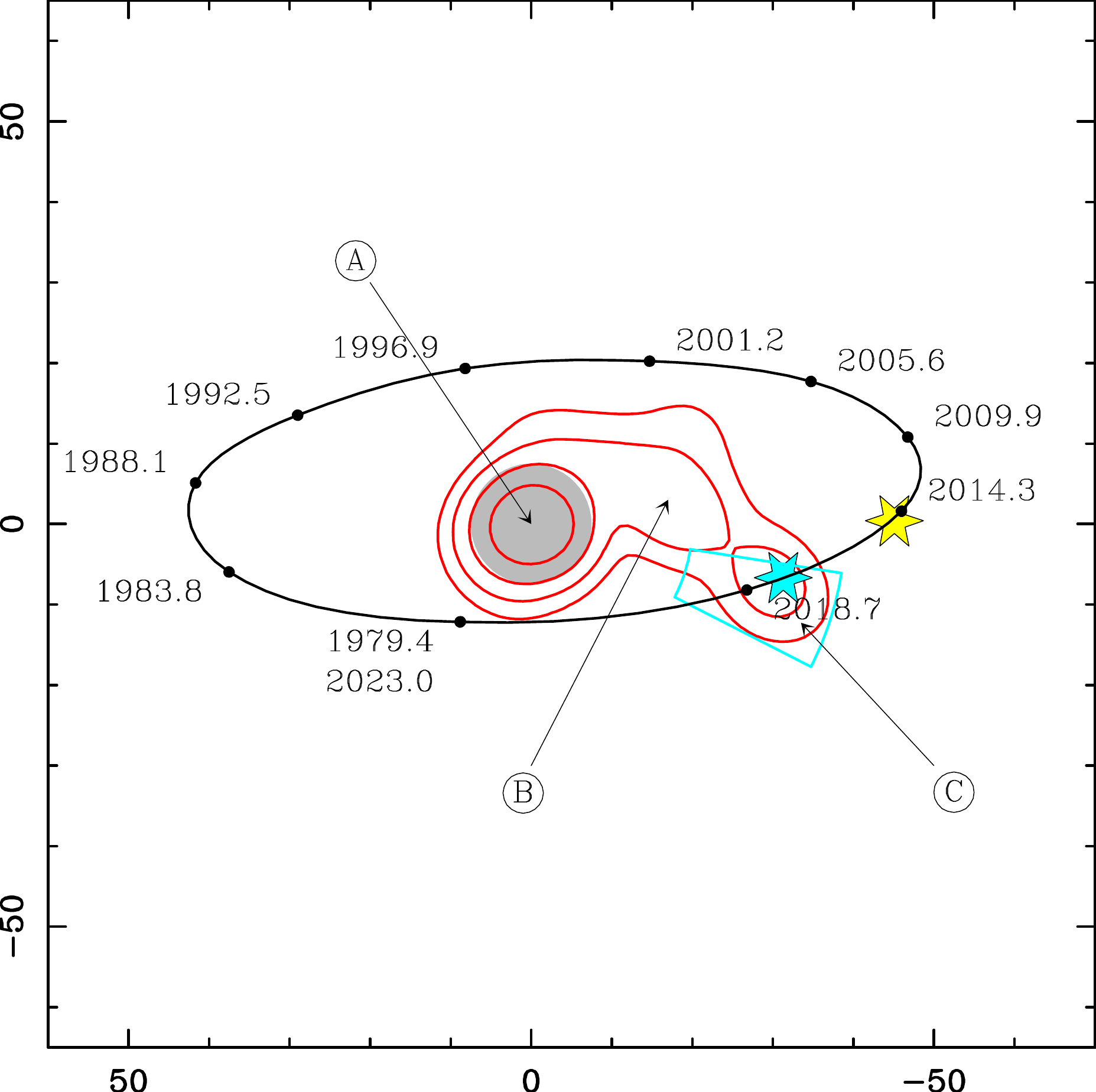}
      \caption{Relative movement of both stars (mas), according to the
        new orbital parameters. The AGB star is represented by the
        gray circle (with a diameter of 15 mas), the position of the
        WD in \citet[][epoch 2014.9]{schmid17} is indicated by the yellow asterisk,
        and the position of the WD derived from the new orbital
        parameters for the epoch of our observations (2017.9) is indicated by the
        blue asterisk. The uncertainties are represented by the blue
        contour. Small black dots show the position of the WD at
        orbital phases of 0.1 to 1.0 by 0.1 (the corresponding dates are also
        indicated).  Our high-resolution continuum map is again shown
        (red contours) with the A, B, and C components; see App.\ A.}
      \label{}
\end{figure}

The determination of the orbital movements is fundamental to comparing the
positions of the stars with the ALMA maps. Since the astrometric
measurements by \citet[][VLT/SPHERE-ZIMPOL observations at epoch
  2014.9]{schmid17}, who relatively placed both stars with a
reasonable accuracy, we can expect a small but noticeable change in
the relative positions for 2017.9.  The best determination of the
orbital parameters of the R Aqr system is that by \cite{gromik09}.
These authors derived the spectroscopic orbit from radial velocity
measurements of the Mira, and used the resolved VLA observation of SiO
masers and continuum emission at 7 mm by \cite{hollis97} to constrain
the major axis and $\Omega$, the orientation of the line of nodes on
sky. Unfortunately, both results by \cite{hollis97} and
\cite{schmid17} are not compatible: the results by Hollis et al.\ imply a
large semimajor axis, $a$ $\sim$ 125 mas, that is 2.7 times larger than the
value resulting from data by Schmid et al.\ ($a$ $\sim$ 47 mas),
although both data predict similar values for $\Omega$, $\sim$
90$\degr$ and 93\degp 5, respectively; we note that there was an
  error in the original number in \cite{gromik09}. The most likely
explanation is that the claimed position for the WD in \cite{hollis97}
is in fact an emission blow in the north jet. Indeed, H$\alpha$ maps by
\cite{schmid17} show bright jet knots near the central source. It is
reasonable to assume that the WD at the epoch of the
VLA observations by \cite{hollis97} was located somewhere between the
two spots they detected, i.e., north of the Mira; the present stellar
positions then indicate retrograde (clockwise) orbital motion.

In summary, we have recomputed the orbital parameters of the system in
order to predict the relative position of the two stars at the epoch of
our ALMA data, using the same radial velocity measurements as in
\cite{gromik09} and the relative position of the stars in
\cite{schmid17}.  The new parameters are the same as in
\cite{gromik09}, except for $i$ (the inclination of the orbit
w.r.t.\ the plane of the sky), which is now 110$^\circ$
(clockwise movement), $\Omega$, which is now 93\degp 5, and
$a$, which is now 47 mas. The resulting new apparent orbit of the
secondary and the prediction for the relative position of the two
stars are shown in Fig.\ B1. We find that the distance between R\,Aqr B (the
WD companion) and R\,Aqr A (the Mira primary) is
32$^{+7}_{-12}$ mas, at PA --102$^{+3}_{-15}$; in our main discussion,
the uncertainties are summarized by a single value of $\sim$ $\pm$7
mas.

In a separate forthcoming paper we plan to discuss in detail the
orbital parameters and improve their determination using data by
\cite{schmid17}, our ALMA measurements, and additional radial
velocities from recent SiO maser data.

\end{document}